\documentclass[a4paper,11pt]{article}
\pdfoutput=1 

\usepackage{jinstpub} 

\usepackage{lineno}
\usepackage{siunitx}
\usepackage{textcomp}
\usepackage{subcaption}
\DeclareSIUnit\bar{bar}
\usepackage{multirow}

\title{First mechanical realization of a tunable dielectric haloscope for the MADMAX axion search experiment}

\abstract{MADMAX, a future experiment to search for axion dark matter, is based on a novel detection concept called the dielectric haloscope. It consists of a booster composed of several dielectric disks positioned with \SI{}{\micro\metre} precision.
A prototype composed of one movable disk was built to demonstrate the mechanical feasibility of such a booster in the challenging environment of the experiment: high magnetic field to convert the axions into photons and cryogenic temperature to reduce the thermal noise. It was tested both inside a strong magnetic field up to \SI{1.6}{\tesla} and at cryogenic temperatures down to \SI{35}{\kelvin}. The measurements of the velocity and positioning accuracy of the disk are shown and are found to match the MADMAX requirements.}


\collaboration{MADMAX collaboration}

\author[a]{B.~Ary Dos Santos Garcia}
\author[a]{D.~Bergermann}
\author[c]{A.~Caldwell}
\author[f]{V.~Dabhi}
\author[f]{C.~Diaconu}
\author[c]{J.~Diehl}
\author[c]{G.~Dvali}
\author[e]{J.~Egge}
\author[e]{M.~Ekmedzic}
\author[f]{F.~Gallo}
\author[e]{E.~Garutti}
\author[b]{S.~Heyminck}
\author[f]{F.~Hubaut}
\author[c]{A.~Ivanov}
\author[g]{J.~Jochum}
\author[f]{P.~Karst}
\author[b]{M.~Kramer}
\author[c]{D.~Kreikemeyer-Lorenzo}
\author[e]{C.~Krieger}
\author[d]{D.~Leppla-Weber}
\author[d]{A.~Lindner}
\author[c]{J.~Maldonado}
\author[c]{B.~Majorovits}
\author[e]{S.~Martens}
\author[d]{A.~Martini}
\author[a]{E.~{\"O}z}
\author[f]{P.~Pralavorio}
\author[c]{G.~Raffelt}
\author[h]{J.~Redondo}
\author[d]{A.~Ringwald}
\author[f]{S.~Roset}
\author[d]{J.~Schaffran}
\author[a]{A.~Schmidt}
\author[c]{F.~Steffen}
\author[g]{C.~Strandhagen}
\author[g]{I.~Usherov}
\author[a]{H.~Wang}
\author[b]{G.~Wieching}
\affiliation[a]{III. Physikalisches Institut A,  RWTH Aachen University, Aachen, Germany}
\affiliation[b]{Max-Planck-Institut f{\"ur} Radioastronomie, Bonn, Germany}
\affiliation[c]{Max-Planck-Institut f{\"ur} Physik, Garching, Germany}
\affiliation[d]{Deutsches Elektronen-Synchrotron DESY, Hamburg, Germany}
\affiliation[e]{Universität Hamburg, Hamburg, Germany}
\affiliation[f]{Aix Marseille Univ, CNRS/IN2P3, CPPM, Marseille, France}
\affiliation[g]{Physikalisches Institut, Eberhard Karls Universit{\"a}t T{\"u}bingen, T{\"u}bingen, Germany}
\affiliation[h]{Universidad Zaragoza, Zaragoza, Spain}

\emailAdd{christoph.krieger@uni-hamburg.de}
\emailAdd{pralavor@cppm.in2p3.fr}

\begin{document}
\maketitle
\flushbottom

\section{Introduction}

The MAgnetized Disk and Mirror Axion eXperiment (MADMAX)~\cite{brun2019} is a future experiment to search for dark matter axions with a mass around \SI{100}{\micro\eV}. This mass, favored when the breaking of the Peccei-Quinn symmetry occurs after an early cosmic inflation~\cite{buschmann2022}, is presently unexplored and challenging to access experimentally. One of the complications lies in the fact that the dark matter axion mass is unknown and needs to be scanned at low temperature ($<\SI{10}{\kelvin}$) and under a high magnetic field ($\gg\SI{1}{\tesla}$). 

To perform such a search, MADMAX is developing a novel concept called the dielectric haloscope~\cite{caldwell2017}. The central part of the detector is a booster, i.e. an arrangement of a metallic mirror and disks of O(\SI{1}{\metre}) diameter made of material with a high dielectric constant $\varepsilon$ as well as low dielectric losses such as sapphire or lanthanum aluminate. The axion to photon conversion occurs inside a strong magnetic field at the boundaries between media with different  $\varepsilon$. The power generated in the booster is directed towards a mirror at one end and ultimately collected by a low-noise receiver system at the other end. When the distance between the disks roughly corresponds to half of the axion wavelength, i.e. of the order of a cm for \SI{100}{\micro\eV} axion~\cite{Millar:2016cjp}, the booster becomes resonant for the converted photons. By re-arranging regularly the distance between the disks, the axion mass can then be scanned. To reduce the thermal noise, the booster will be housed inside a cryostat working at liquid helium temperature. With such a setup, the very low power emitted from dark matter axions, of the order of \SI{e-27}{\watt}/m$^2$, can be boosted to \SI{e-22}{\watt}/m$^2$ in a $\SI{10}{\tesla}$ field, assuming axions make up all of the dark matter~\cite{brun2019}. 

The MADMAX booster will ultimately be placed in a $\sim\SI{9}{\tesla}$ dipole magnet to be located in the iron yoke of the former H1 experiment~\cite{H1:1996prr} at DESY in Hamburg (Germany). It is presently in a prototyping phase and the purpose of this article is to describe the mechanical performance obtained with one prototype. The latter consists of one movable \SI{200}{\milli\metre} diameter sapphire disk plus a mirror and is called open booster 200 (OB200). The disk is moved by three piezoelectric motors~\cite{PiezoProto} mounted on rails. OB200 was designed to demonstrate the mechanical feasibility of the MADMAX booster and measure the performance of the disk movement at cryogenic temperatures as well as inside a strong magnetic field. The MADMAX requirements in terms of velocity and absolute disk position accuracy are at least \SI[per-mode=symbol]{10}{\micro\metre\per\second} and less than \SI{25}{\micro\metre} min-to-max~\cite{MADMAX:2021lxf}, respectively. The minimal velocity is driven by the requirement of the disk positioning time not being the dominant contribution when reconfiguring the booster during the scan. Further requirements are that the three motors should not be more than \SI{100}{\micro\metre} apart during the movement, to avoid stressing the disk, and that they remain within \SI{1}{\micro\metre} when reaching their final position.

The article is organized as follows. Sections~\ref{sec:Mechanics} and~\ref{sec:Motors} describe the OB200 mechanics and the precision actuation system including the integrated interferometer system used to position the disks, respectively. The test setups and the results performed at room temperature with or without magnetic field and at cryogenic temperature are reported in section~\ref{sec:Results}. Section~\ref{sec:Conclusions} is devoted to conclusions.

\section{Description of prototype mechanics}
\label{sec:Mechanics}

\begin{figure}[htbp]
    \centering
    \includegraphics[height=4.4cm]{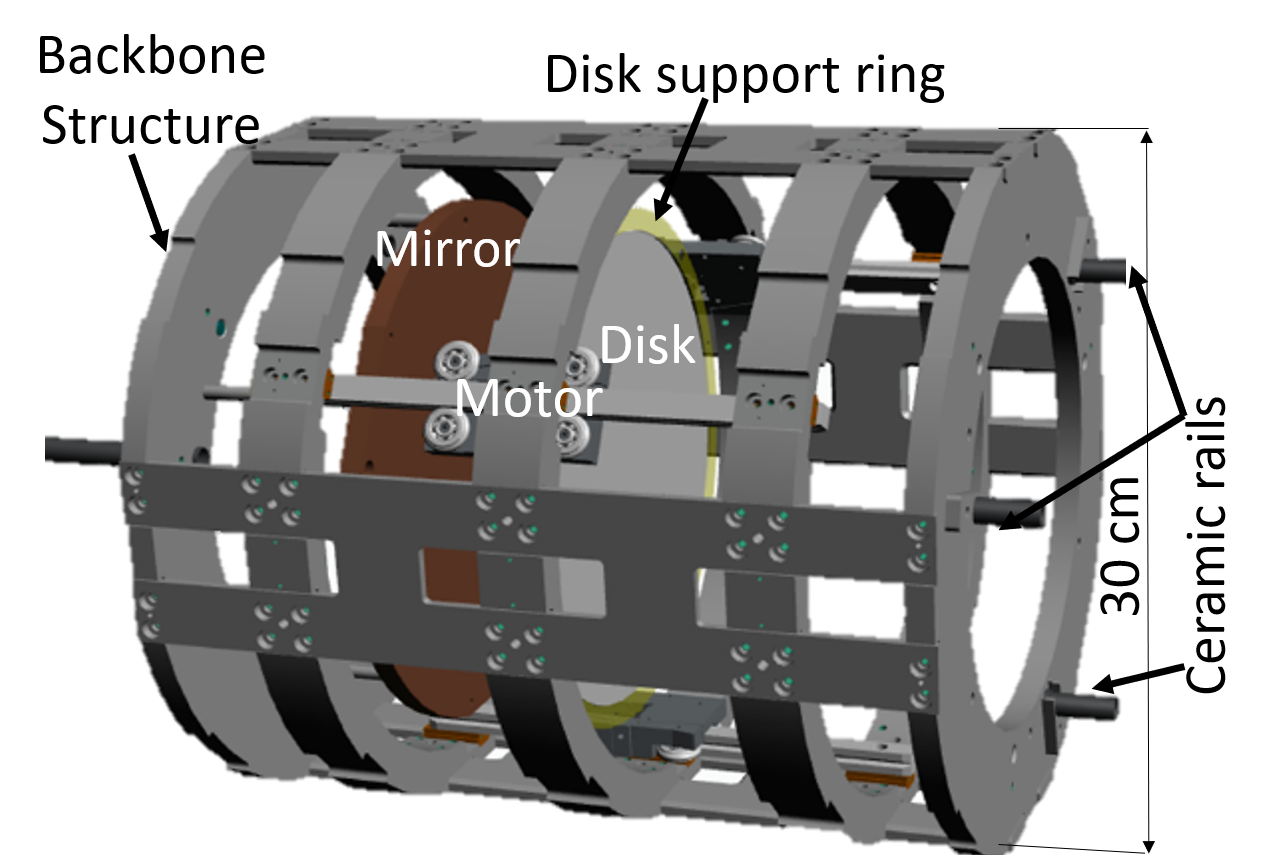} 
    \includegraphics[height=4.4cm]{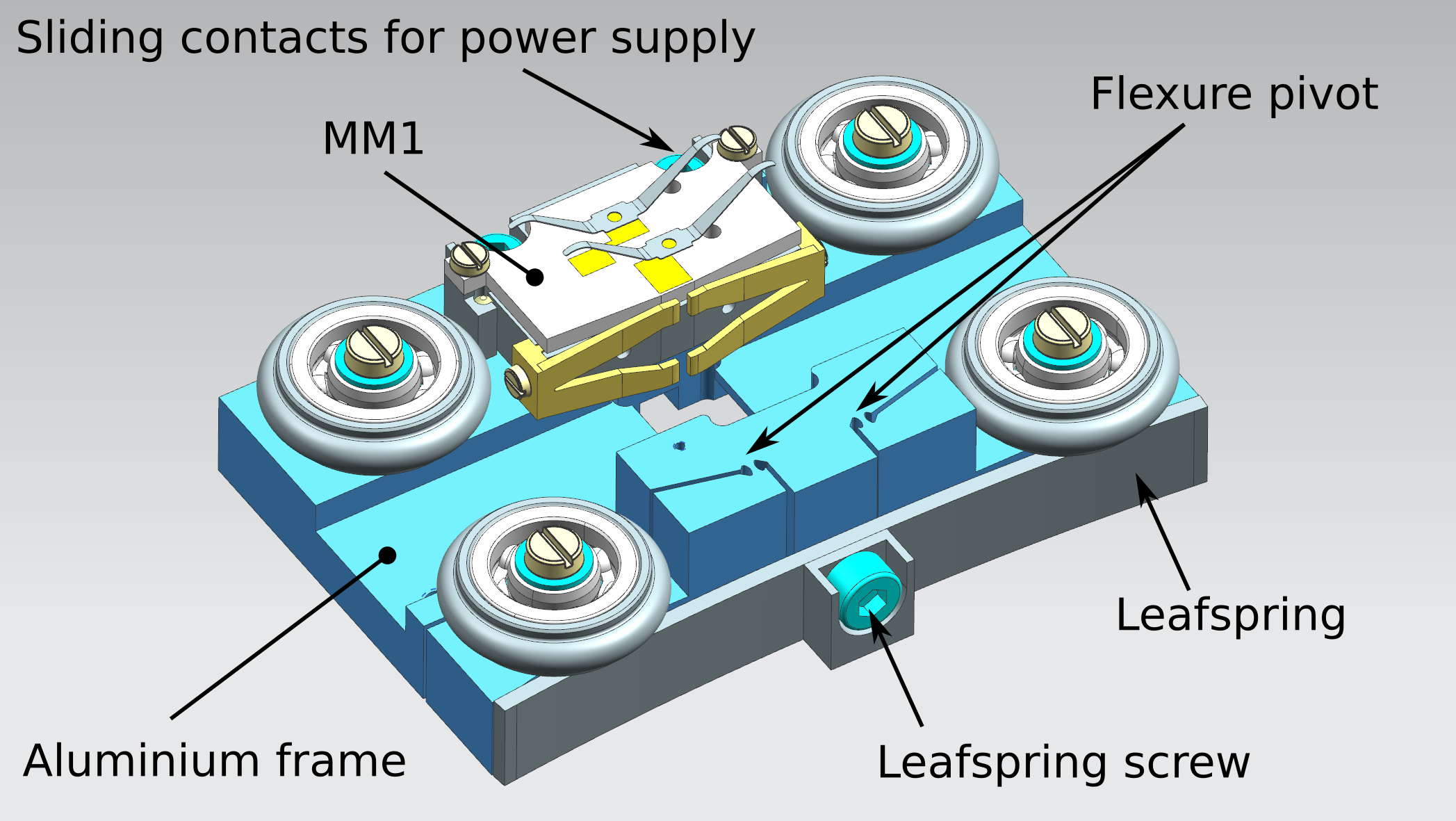}
    \caption{Left: OB200 design concept with one disk and one mirror. Right: zoom on the carriage with the piezo motor, taken from~\cite{PiezoProto}.}
    \label{fig:P200_sketch}
\end{figure}

\begin{figure}
    \centering
    \includegraphics[width=0.9\linewidth]{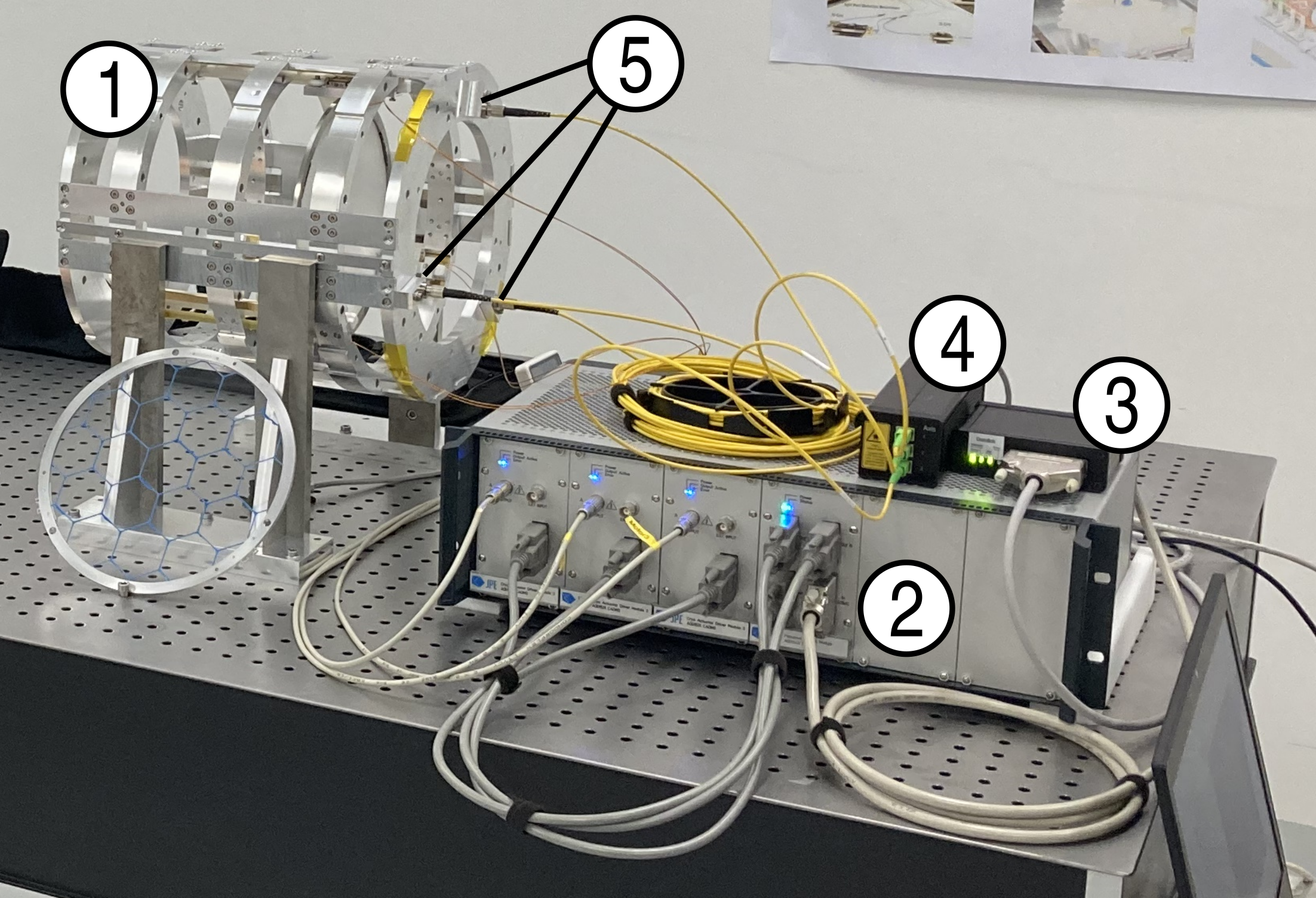}
    \caption{Photograph showing the OB200 setup: backbone structure with disk drive system and single sapphire disk mounted in disk support ring (1), motor and disk drive controllers including the feedback module (2), FPGA board (3), and laser interferometer (4). The optical sensor heads (5) mounted to the backbone structure are connected via yellow optical fibres to the Attocube IDS which monitors the movement of retroreflectors installed on the disk support ring. The position information is fed into the disk drive controller which steers the three piezo motors via the respective motor controllers.}
    \label{fig:P200Photo}
\end{figure}

OB200 is a mechanical structure in the form of a \SI{500}{\milli\metre} long cylinder with a diameter of \SI{300}{\milli\metre} (Figure~\ref{fig:P200_sketch} left). It is equipped with a \SI{200}{\milli\metre} diameter disk moved with three motors (Figure~\ref{fig:P200_sketch} right) positioned at \SI{120}{\degree} from each others in the azimuthal direction and sliding on three parallel rails. The rails are inserted into a precisely machined mechanical structure which supports the disk drive system on which a ring, holding the disk, is mounted. A picture of the complete setup is shown in Fig.~\ref{fig:P200Photo}. Additionally, a flat metallic mirror (made from copper with a thickness of \SI{10}{\milli\metre} and machined to a flatness better than \SI{10}{\micro\metre}) can be mounted to the mechanical structure to make it a fully functional minimalistic booster.

\subsection{Backbone structure}
\label{sec:backbone}

The backbone structure is made from structural aluminium and consists of three structure rings, which feature various interfaces for the disk drive system. They are connected via four bridge pieces and two end rings. The latter provide the mechanical interface to the laser interferometer used to monitor the disk position. The overall design of the backbone structure aims at a symmetrical shrinkage during cool-down to cryogenic temperatures to keep the alignment between the laser interferometer, the disk (support ring) and the motors intact. The components of the backbone structure are precisely machined with an accuracy of better than \SI{10}{\micro\metre}. 
During assembly, the parts are aligned using precision ground cylinder pins before being bolted together using stainless steel hardware. The cylinder pins are made from hardened steel and therefore are removed after assembly to make the backbone structure compatible with usage inside strong magnetic fields. 

\subsection{Disk drive system}
\label{sec:driveRail}

The system used to move the disk inside the backbone structure consists of three ceramic rails made from zirconium oxide. Three carriages featuring ceramic ball bearings and wheels are positioned on these rails. The carriages are moved by the piezoelectric motors.

The ceramic rails need to be installed parallel to each other to allow for moving a single disk along the direction of the rails with the three piezoelectric motors. At the same time the rails need to be perpendicular to the interferometer rings of the backbone structure to ease the alignment of the laser interferometer which uses retro reflectors, placed on the disk support ring, to measure the displacement of the carriages. 

\subsection{Disk support ring}
\label{sec:ring}

\begin{figure}[htbp]
    \centering
    \includegraphics[width=0.85\linewidth]{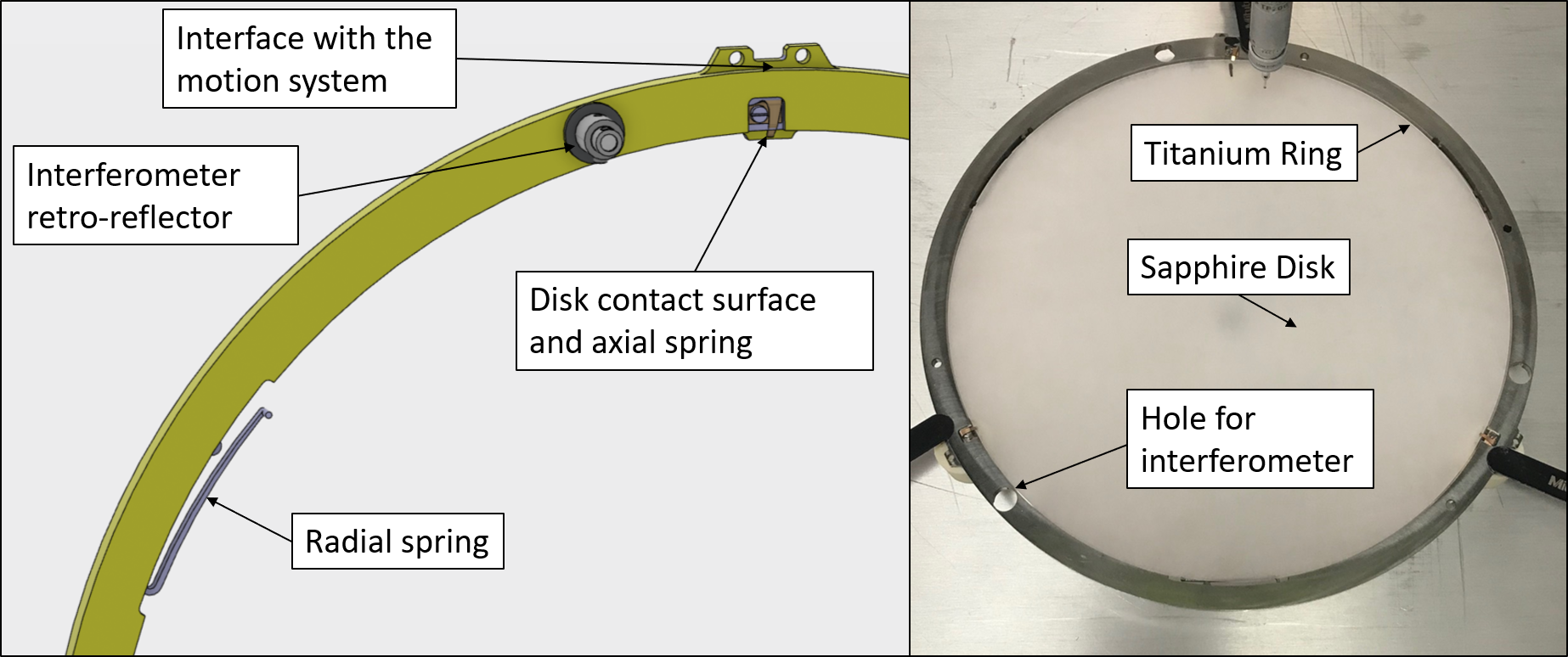}
    \caption{Design (left) and photograph (right) of the disk support ring. The axial and radial springs, the holes for the interferometer mirrors and the extensions for interfacing the motor carriages are indicated (see text for more details).}
    \label{fig:ring_sketch}
\end{figure}

In order to hold the OB200 sapphire disk and interface it with the three motor carriages, a support ring has been designed and built, as shown in Figure~\ref{fig:ring_sketch}. It is made of titanium, to ensure non-magnetism, its inner (outer) diameter is \SI{201}{\milli\metre} (\SI{204}{\milli\metre}) with a thickness of \SI{4}{\milli\metre}. Three small axial springs made of copper-beryllium are screwed on the ring. This allows to clamp the disk on its outer radius against three small contact surfaces. Three radial springs are machined on the ring to ensure the radial centering of the disk and to accommodate the different material expansion coefficients during cool-down. The connection with the three piezo motors is made via three extensions at the ring's outer radius using titanium leaf springs, which are screwed onto the motor carriages. This compensates for the differential shrinking between disk support ring and backbone structure during cool-down. Last but not least, three holes are drilled in the ring, at 120 degrees from each other, to position the retroreflectors used by the interferometer system.

The main challenge for such a thin support ring is that it should not alter the flatness of the disk surfaces to a level of precision better than \SI{10}{\micro\metre}. To meet this requirement, a complex fabrication procedure has been developed. After rough cutting operations and heat treatment for stress relief of the raw material, the ring is machined on a bed of resin. This allows to minimize mechanical deformations during the manufacturing of the ring. The final operation is the machining of the three radial springs by wire-cut Electrical Discharge Machining.

\subsection{Disk metrology}
\label{sec:metro}

\begin{figure}[htbp]
    \centering
    \includegraphics[width=0.99\linewidth]{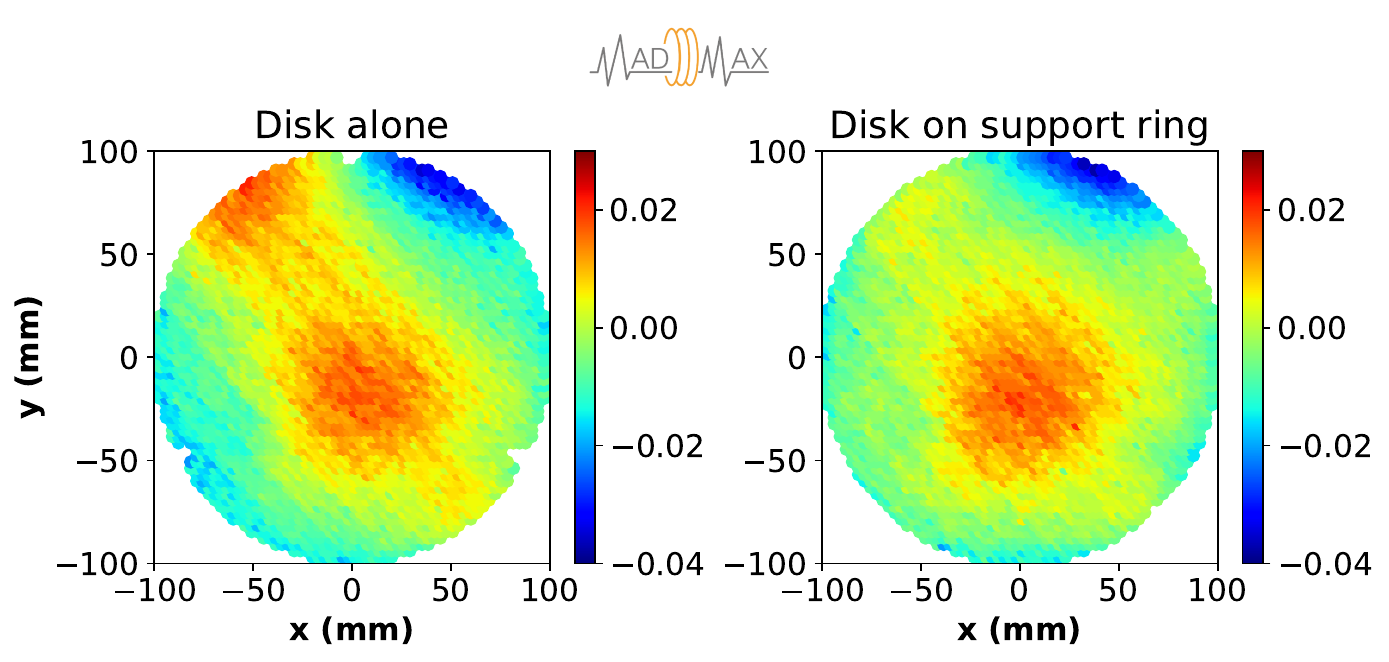}
    \caption{Precise measurement in vertical position of one disk surface flatness before (left) and after (right) its mounting on the support ring. The color scale unit is in \SI{}{\milli\metre}.}
    \label{fig:ring_meas}
\end{figure}

A precise metrology of the ring has been performed with a three-dimensional positioning scanning Machine\footnote{Tri Mesure Tempo MCA10}. It shows a flatness of the plane defined by the three contact surfaces with the disk at the level of \SI{3}{\micro\metre} (min-max). Measurements of the disk in vertical position have also been performed, before and after it is mounted on the ring. The RMS of the measurements of the disk alone is below \SI{10}{\micro\metre} for each side. This is still the case after mounting the disk on the ring. Its geometry is only slightly modified at the edges, as seen in Figure~\ref{fig:ring_meas}. 

\section{Precision actuation system}
\label{sec:Motors}

To position the disk with high precision, an actuation system is used. It consists of three piezoelectric motors that are running on individual ceramic rails, with individual motor drivers/controllers (Cryo Actuator Driver Module 3), a control module (Flexdrive Control Module) allowing the coordinated driving of the three piezo motors to a set target position, and an interferometer used to monitor the individual motor positions. The interferometer is linked electronically to the control module to provide absolute disk positions. The complete system is shown in Fig.~\ref{fig:P200Photo}.

\subsection{Piezoelectric motors}
\label{sec:piezo}

The piezo motors use a new design, based on the stick-slip technology, developed by the MADMAX collaboration in cooperation with
the company JPE\footnote{JPE, Aziëlaan 12, 6199 AG Maastricht-Airport, The Netherlands, www.jpe-innovations.com}. It was tested successfully at cryogenic temperature (around \SI{4}{\kelvin}) and under high magnetic field conditions (\SI{5.3}{\tesla})~\cite{PiezoProto}. It ensures high precision (better than \SI{10}{\micro\metre}) along a long stroke of typically tens of centimetres. The voltage signal to drive the piezoelectric motor is provided via sliding contacts on a PCB with contact strips. The voltage signal is generated by a dedicated driver module for each motor. The motors are mounted onto small carriages running on the ceramic rails using ceramic wheels and ball bearings. The required mechanical preload is provided via flexing features in the carriage in combination with a titanium leaf spring. 

\subsection{Driving system}
\label{sec:drive}

As the piezoelectric motors are exploiting the stick-slip principle for movement along the ceramic rail, the step size varies between the motors requiring a dedicated driving system with a position feedback signal for a coordinated movement of the disk. The step sizes are temperature-dependent, typically O(\SI{10}{\micro\metre}) at room temperature and O(\SI{1}{\micro\metre}) at cryogenic temperature. The movement of the disk is performed by motor drivers, configured for external control and connected to the module which retrieves the position information from the laser interferometer via a synchronous serial interface (SSI). The motor drivers and the feedback module are mounted in a single cabinet as shown in Fig.~\ref{fig:P200Photo}. 

The control module operates the primary motor (master) with a fixed frequency of typically \SI{50}{\hertz}, and thus velocity, while the frequency of the secondary motors (slave) is varied in the range \SIrange{0}{70}{\hertz} to match the velocity of the primary motor. Typically, all motors are operated with Relative Stepsize (RSS) set to \SI{100}{\percent}, meaning with the maximum stepsize at the given temperature. The motors are stopped once each of them reaches the target position within a tolerance value which should be below \SI{10}{\micro\metre}. Additionally, a maximum allowed distance between the three motors during movement is set (master-slave distance, abbreviated as MSD). If the MSD value is exceeded, the movement of the disk is either stopped or in case the primary motor is lagging behind the secondary motors, the secondary motors halt until the primary motor draws closer.

In the following, a movement of the disk via the disk drive system by a certain distance is called a jump.

\subsection{Interferometer for closed-loop control}
\label{sec:interf}

As the position measurement device has to work reliably at cryogenic temperatures, a commercially available laser interferometer is used to provide the required input for the control module. The interferometer used is an Attocube\footnote{attocube systems AG, Eglfinger Weg 2, 85540 Haar, Germany, www.attocube.com} IDS 3010 in combination with optical sensor heads and retro-reflectors rated for operation in vacuum and at cryogenic temperatures. The sensor heads are mounted to the backbone structure, and the retro-reflectors are fixed either directly to the motor carriages (for measurements without a disk installed) or to the disk support ring. The Attocube IDS 3010 can measure any displacement from the initial position at start-up of the interferometer measurement with a sub-nanometre accuracy, which is more than sufficient to reach a positioning accuracy of the motors/disk in the micrometre range. To achieve a reliable measurement, the retro-reflectors need to be aligned with the optical sensor heads such that over the whole travel range a sufficient contrast in the interferometer signal is maintained. Here, the usage of retro-reflectors instead of flat optical mirrors eases the alignment. The interferometer is connected to the sensor heads via optical fibres either directly (as shown in Fig.~\ref{fig:P200Photo}) or with an off-the-shelf fibre-optical vacuum feedthrough for operation inside a cryostat.

The measured displacements are transmitted by the Attocube IDS 3010 via a High Speed Serial Link (HSSL). To adapt the data format to the SSI used by the control module, a custom-programmed FPGA board is used. The output of the Attocube IDS 3010 is configured in a way to allow for a maximum displacement of~\SI{4.4}{\metre} with a resolution of about \SI{33}{\nano\metre} given the data format used by the control module. The values in the FPGA buffer are updated with a rate of several \SI{}{\kilo\hertz}. 

\section{Disk velocity and positioning accuracy measurements}
\label{sec:Results}

To test OB200, and especially the disk drive system inside a strong magnetic field and at cryogenic temperatures, two different setups are used. Both setups can be used with and without the disk (connected to the support ring) installed. Operation without disk allows for measuring the velocity and step size of the individual motors, and for testing the coordinated movement of the three motors without a disk support ring mechanically linking the three carriages. For measurements without disk the retro-reflectors are mounted to the carriages directly. When the disk is mounted to the carriages, the retro-reflectors are unmounted from the carriages and attached to the foreseen positions on the disk support ring, requiring a realignment of the optical sensor heads with respect to the retro-reflectors.

\subsection{Test setups}
\label{sec:setup}

\begin{figure} [htbp]
 \centering
    \includegraphics{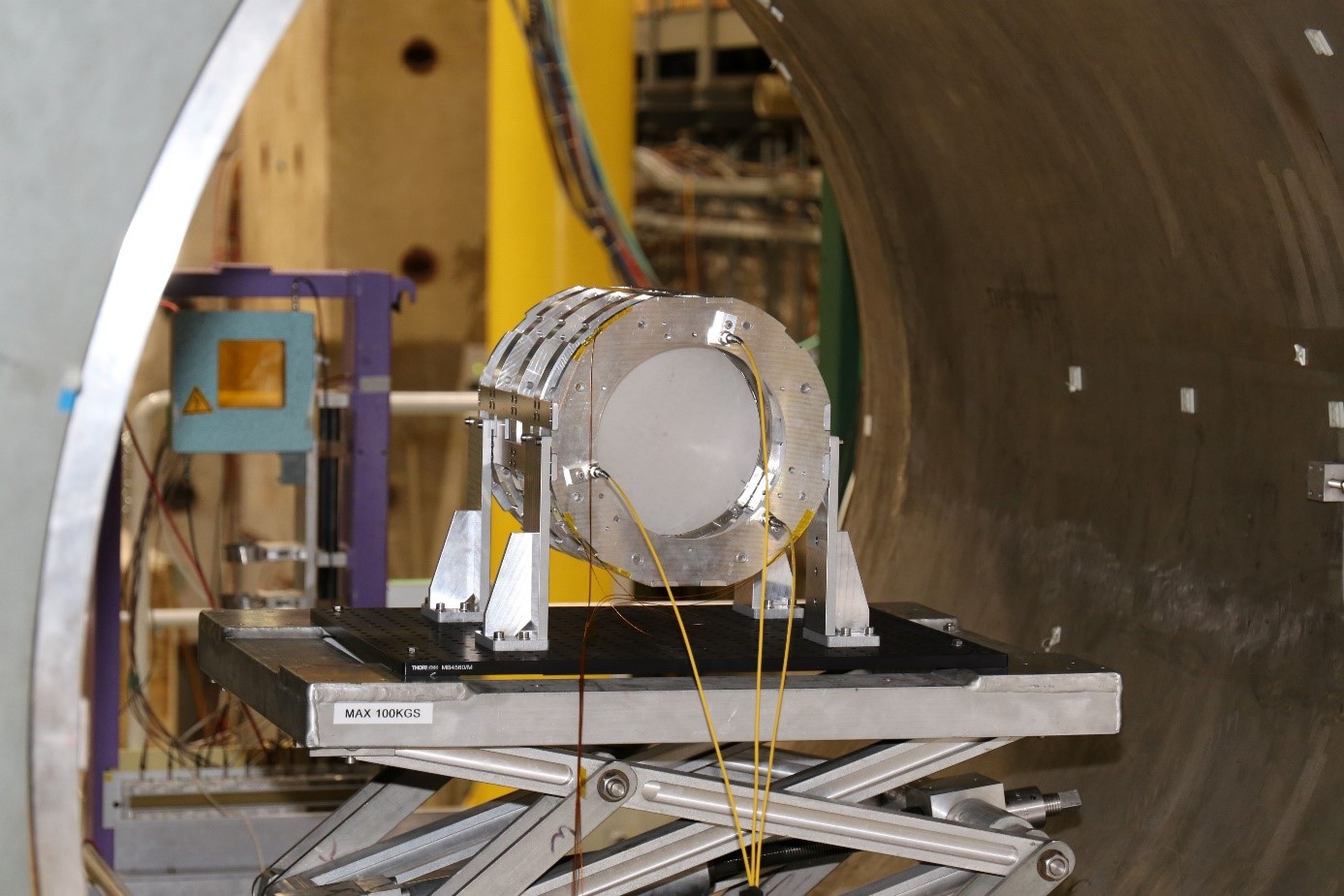}
    \caption{OB200 inside the CERN Morpurgo magnet.}
    \label{fig:setup_morpurgo}
\end{figure}

For the test under high magnetic field, OB200 is installed on a movable trolley which was inserted into the \SI{1.6}{\tesla} magnetic field of the Morpurgo dipole magnet at CERN~\cite{morpurgo}. The trolley allowed for taking the setup outside of the magnetic field for reference measurements without the need of ramping down the magnet. 
A picture of the setup inside the Morpurgo magnet is shown in Fig.~\ref{fig:setup_morpurgo}. Here, the Attocube IDS 3010 is connected to the optical sensor heads directly using the room temperature optical fibres without any additional feedthrough or connectors in between.\\

\begin{figure} [htbp]
 \centering
    \includegraphics[height=6.75cm]{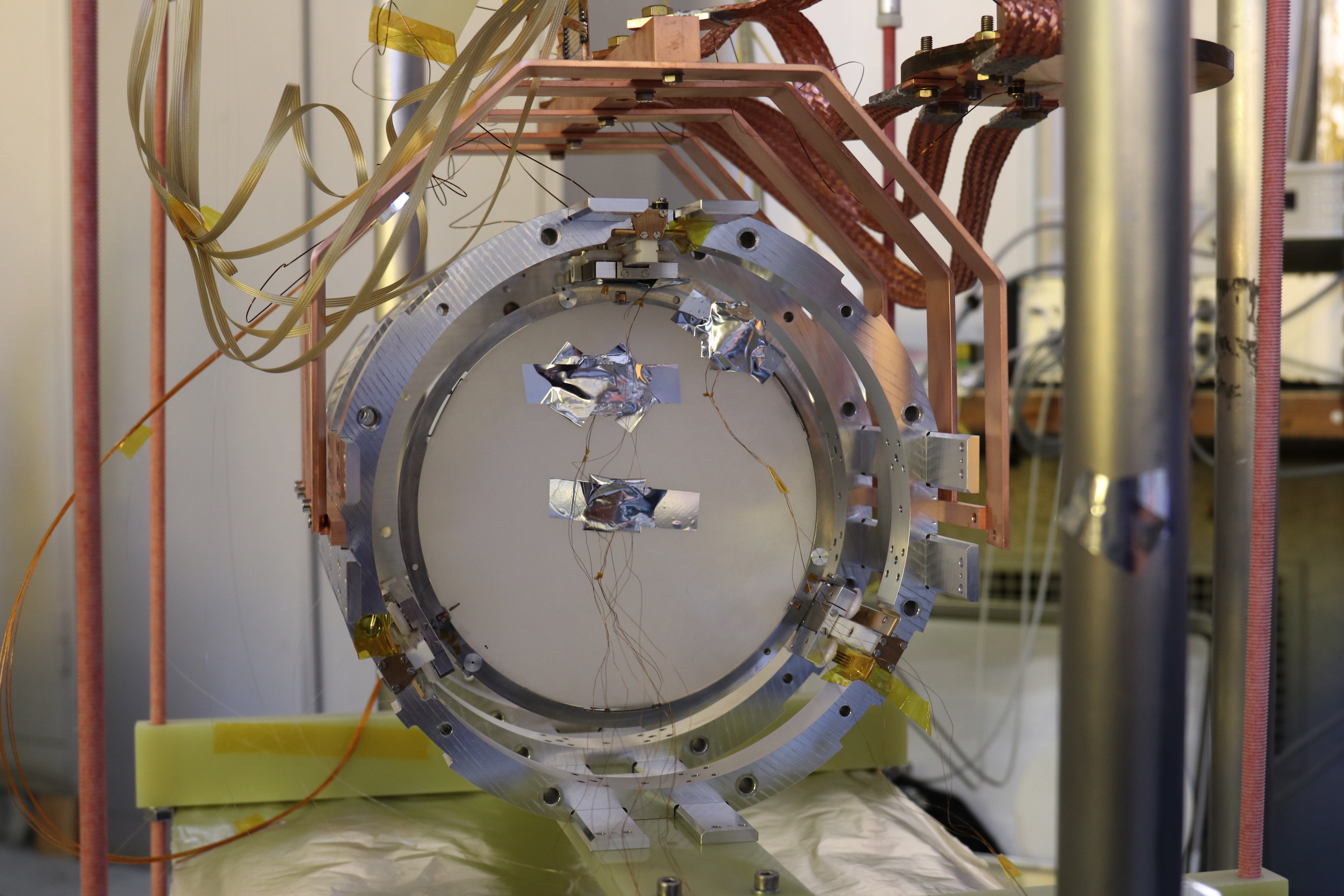}
    \includegraphics[angle=90,height=6.75cm]{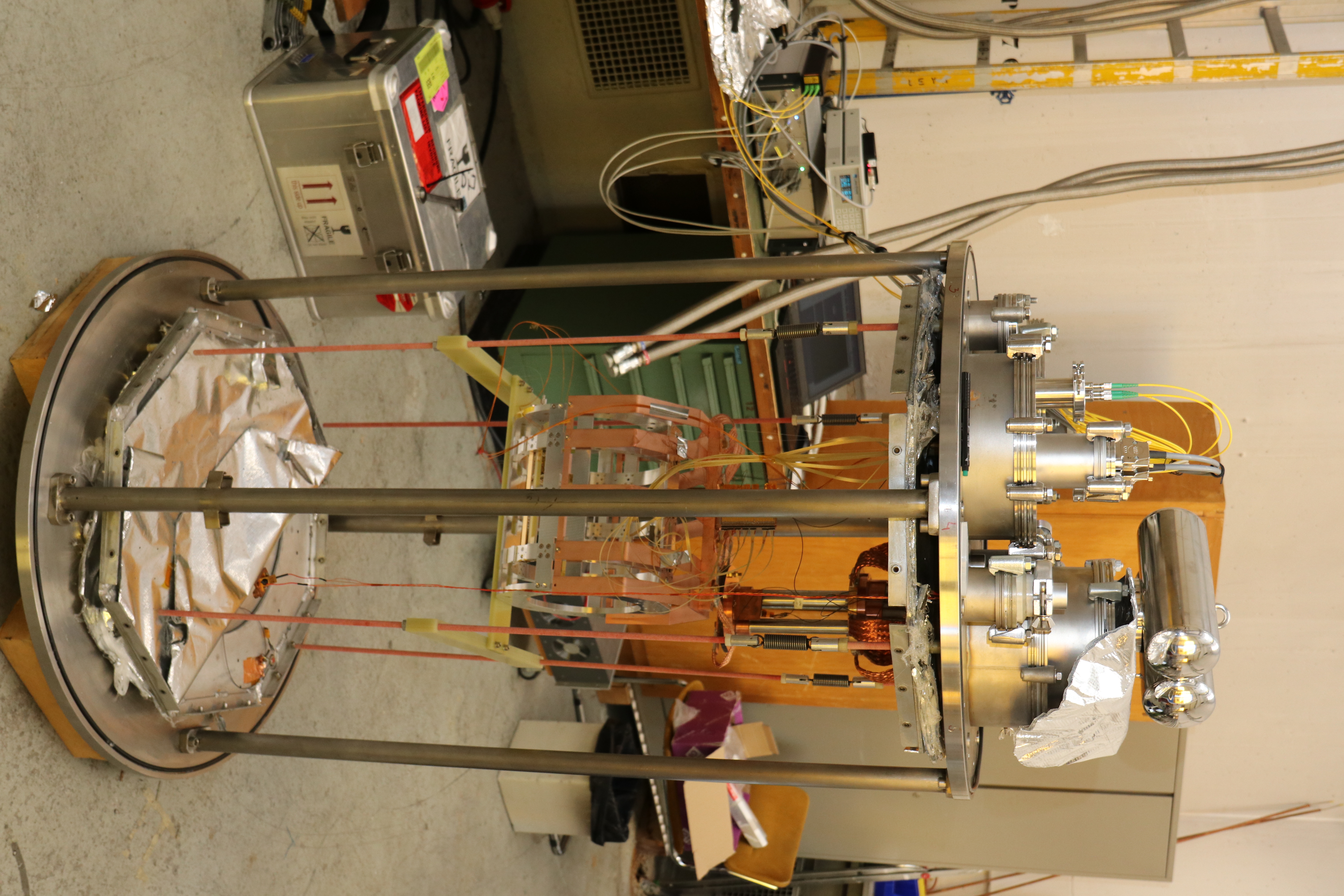}
    \caption{Left: OB200 installed in the cryostat at CERN. The backbone structure is thermally linked to the second stage of a pulse tube cryocooler via copper braids and links. Temperature sensors are mounted to the backbone structure, a motor carriage and the sapphire disk. Right: Cryostat without vacuum shell and thermal shields. OB200 is suspended inside the cryostat via G11 rods and a structure made from G10.}
    \label{fig:setup_cryo}
\end{figure}

To reach cryogenic temperatures, a dry cryostat with large volume on loan from the CERN Magnet Group is used at the CERN Central Cryogenic Laboratory. A picture of the setup inside the cryostat is shown in Fig.~\ref{fig:setup_cryo}. The cryostat is cooled by a single dual-stage pulse tube cryocooler. While the first stage cools the thermal shield, the second stage directly cools the backbone structure of OB200 which is connected via a copper cage and copper braids. Temperature sensors are installed on the backbone structure, the motor carriages (for the tests without disk) or to one motor carriage and the center of the disk as well as on the disk support ring (for the tests with disk installed). OB200 is suspended inside the cryostat´s insulation vacuum on a structure made from G10 and G11. To further minimize the heat radiation to the setup, a thermal screen made from very thin insulation material is placed between OB200 and the thermal shield covered with MLI blankets. For the laser interferometer, a commercially available fibre-optic vacuum feedthrough is used to feed the optical fibres from OB200 to the Attocube IDS 3010 outside the cryostat. For the electric signals of the piezo actuators, a standard D-Sub vacuum feedthrough is used.

\subsection{Results at room temperature with and without magnetic field}
\label{sec:RT}

\begin{figure}[t]
    \centering
    \includegraphics[width=0.485\linewidth]{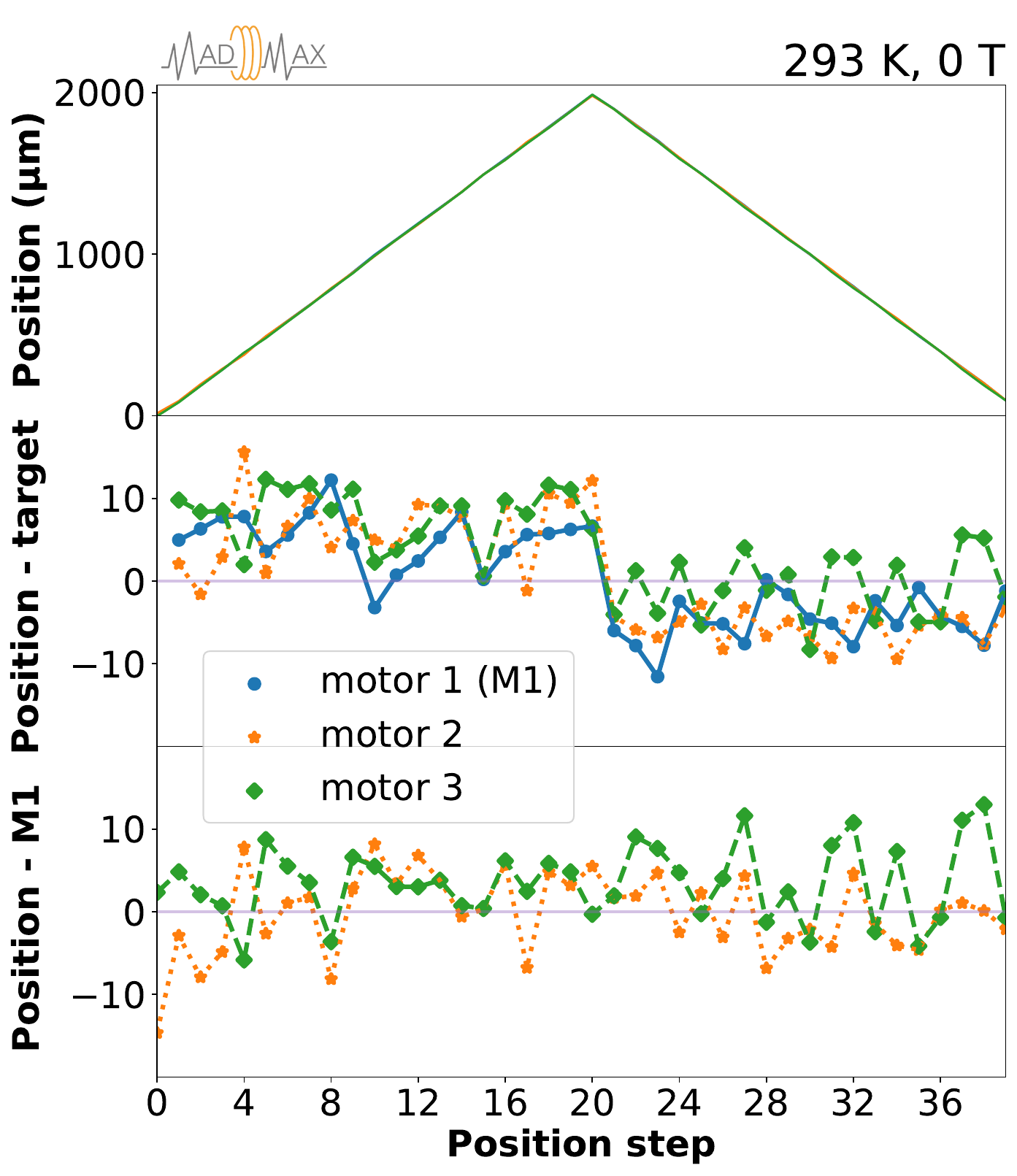}
    \includegraphics[width=0.50\linewidth]{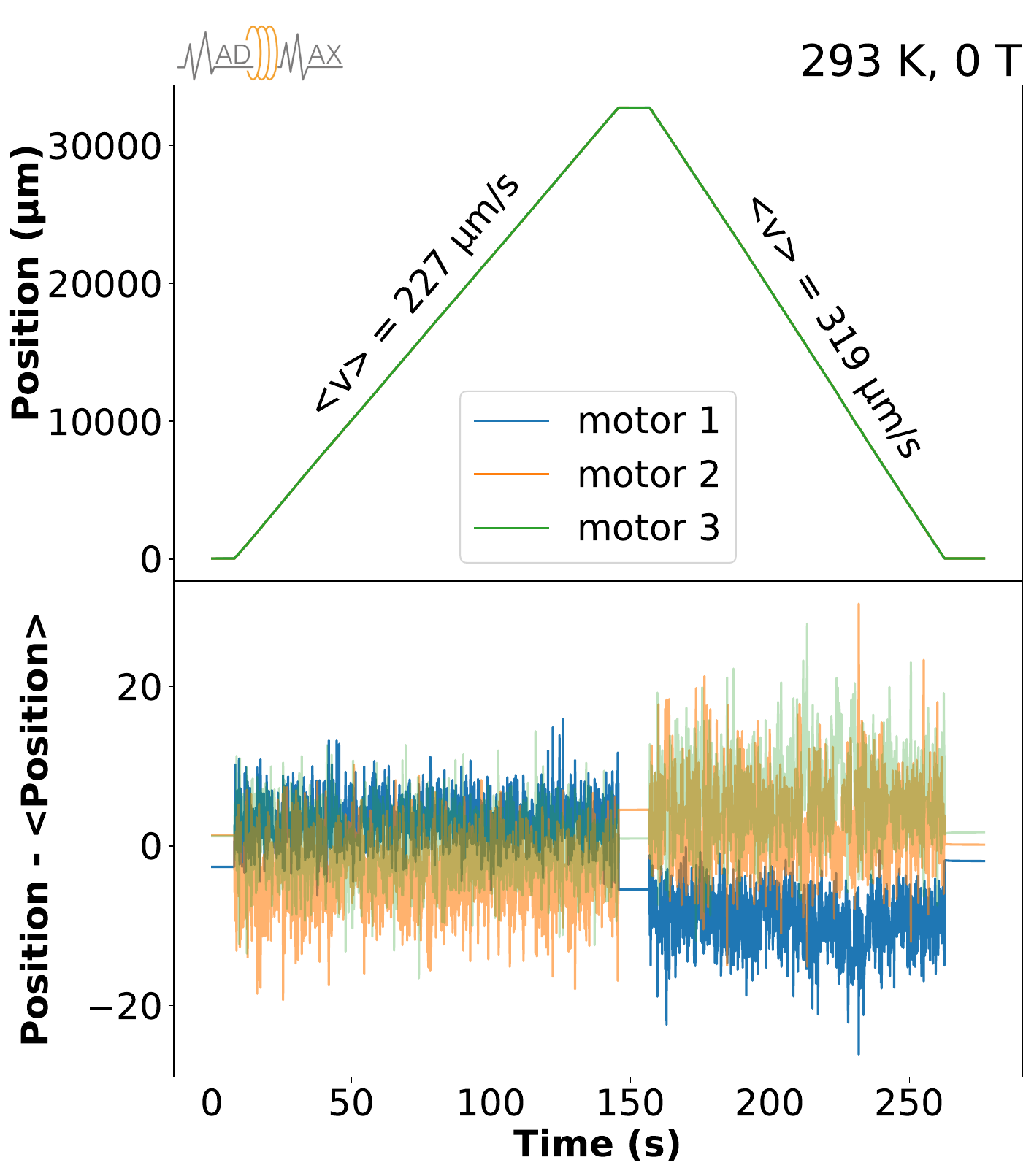}
    \caption{Measurements of OB200 at room temperature with no magnetic field. Left: Positioning accuracy measurements in \SI{}{\micro\metre} using the motor system for a typical back and forth movement. The evolution as a function of the motor position step (reached after each jump of the motor by \SI{100}{\micro\metre}) from the starting point is shown in the top panel. The distance between the positions of the three motors and the target is shown in the central panel and the distance between motors~2 and 3 and motor~1 in the bottom panel. In both cases, the errors on each point are lower than 1~nm (not visible on the plot) and lines are connecting the points to ease the plot readability. Right: Motor displacement in \SI{}{\micro\metre} as a function of time in seconds for a full back and forth movement across OB200 is shown in the top panel. Deviation between each motor and the average position of the three motors as a function of time is shown in the bottom plot. Measurements are taken every \SI{0.11}{\second}.}
    \label{fig:Detail_Pos_RT}
\end{figure}

Figure~\ref{fig:Detail_Pos_RT} top left plot shows the evolution of the motor positions during the test at room temperature under no magnetic field with a jump length of \SI{100}{\micro\metre}. A complete cycle corresponds to a back and forth displacement of $20\times\SI{0.1}{\milli\metre}$ from the original position (\num{20} jumps per direction). The position difference between the three motors and the target is shown in the central plot. All motors can reach the target position after each jump with a precision better than \SI{15}{\micro\metre}, the achieved precision corresponds to the step size of the motors as expected. The difference between motors~2 and 3 and motor~1, which acts as the master, is shown in the bottom plot. The position of motor~1 is within \SI{15}{\micro\metre} of the other motors. The stability of the motor position, once the target position is reached, is always better than \SI{1}{\micro\metre}. Same results are obtained when considering \num{1000} and \SI{2000}{\micro\metre} for the jump length.

Another run with the motor positions being registered every \SI{0.11}{\second} is considered. Figure~\ref{fig:Detail_Pos_RT} right shows the motor position as a function of time (top) and the deviation of each motor from the average position (bottom). Noticeable differences are observed between the back and forth movements: the velocity increases for the reverse movement, reaching more than \SI[per-mode=symbol]{300}{\micro\metre\per\second}. This was already observed during the single motor test~\cite{PiezoProto}. Second, the distance between motors during movement is increased from \SIrange{5}{20}{\micro\metre} -- a direct consequence of the velocity increase, but is still largely below the specifications of \SI{100}{\micro\metre}.

\begin{figure}[t]
    \centering
    \includegraphics[width=0.485\linewidth]{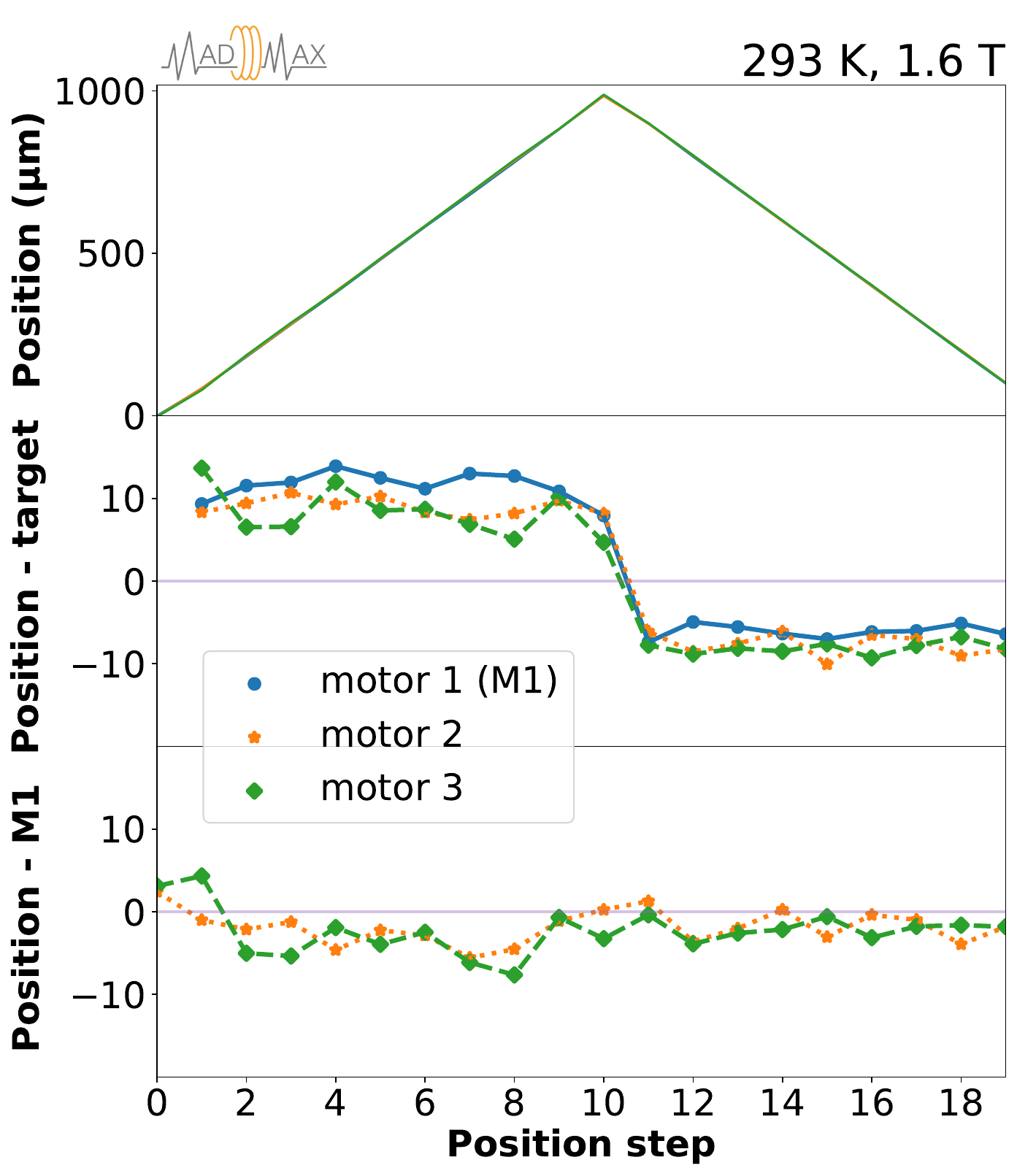}
    \includegraphics[width=0.491\linewidth]{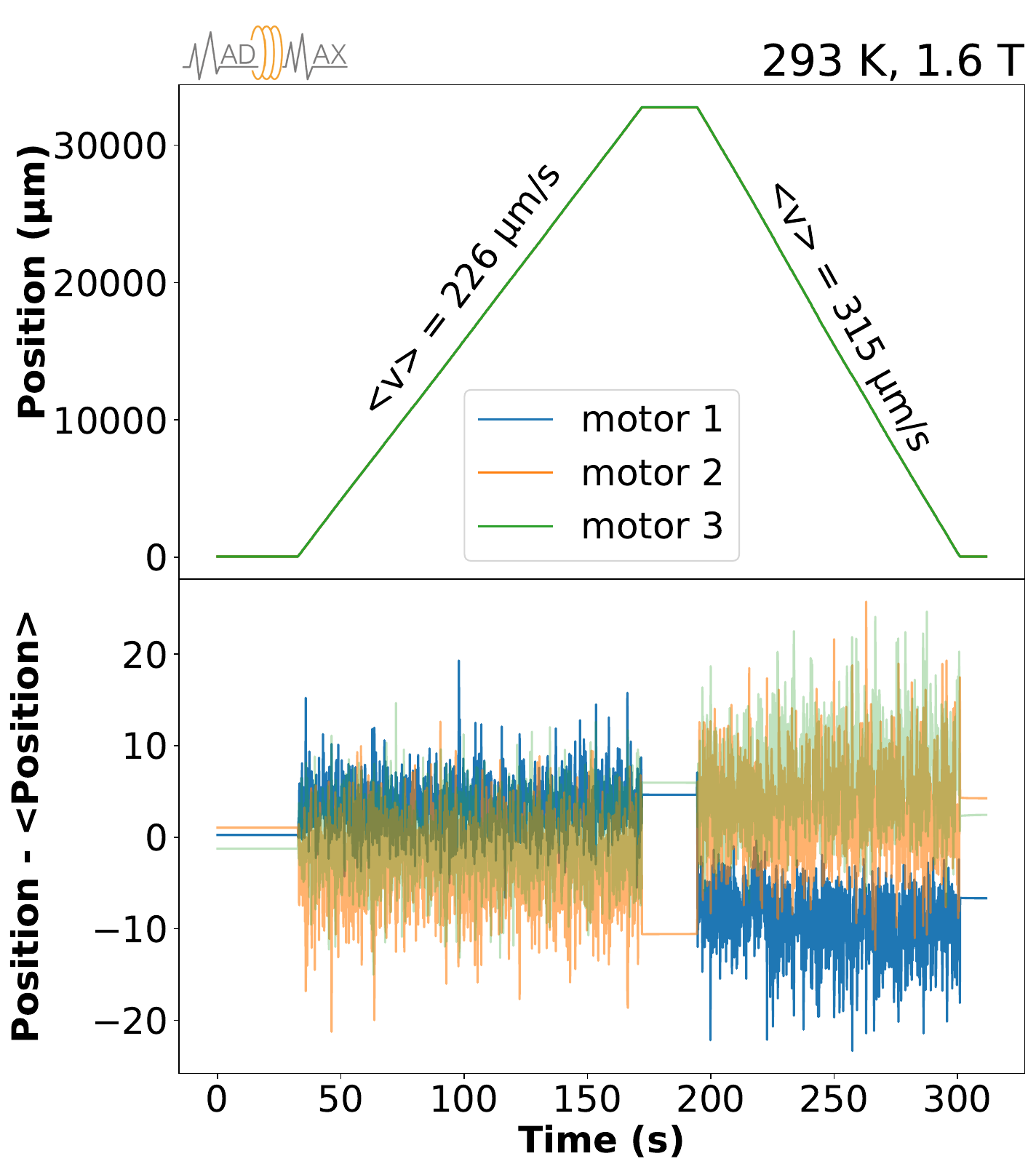}
    \caption{Measurements of OB200 at room temperature under \SI{1.6}{\tesla} magnetic field. Left: Positioning accuracy measurements in \SI{}{\micro\metre} for a typical back and forth movement. The evolution as a function of the motor position step (reached after each jump of the motor by \SI{100}{\micro\metre}) from the starting point is shown in the top panel. The distance between the positions of the three motors and the target is shown in the central panel and the distance between motors~2 and 3 and motor~1 in the bottom panel. In both cases, the errors on each point are lower than 1~nm (not visible on the plot) and lines are connecting the points to ease the plot readability. Right: Motor displacement in \SI{}{\micro\metre} as a function of time in seconds for a full back and forth movement across OB200 is shown in the top plot. Deviation between each motor and the average position of the three motors as a function of time is shown in the bottom plot. Measurements are taken every \SI{0.11}{\second}.}
    \label{fig:Detail_Pos_RT_B}
\end{figure}

Figure~\ref{fig:Detail_Pos_RT_B} shows the same plots at room temperature under \SI{1.6}{\tesla} magnetic field. Figure~\ref{fig:Detail_Pos_RT_B} left shows that all motors can reach the target position after each jump with a precision better than \SI{15}{\micro\metre}, demonstrating that the magnetic field has only marginal impact on the performance of the motor positioning. Similarly, the difference between motors~2 and 3 and motor~1 is always within \SI{5}{\micro\metre} and the stability of the motor position once the final target position is reached is always better than \SI{1}{\micro\metre}. Tests performed with a jump length of \num{10} and \SI{2000}{\micro\metre} produce the same outcome. Figure~\ref{fig:Detail_Pos_RT_B} right shows the motor position as a function of time (top) and the deviation of each motor from the average position (bottom). Compared to Figure~\ref{fig:Detail_Pos_RT}, the velocity is not affected by the \SI{1.6}{\tesla} magnetic field. The distance between motors during movement is still largely below the specifications of \SI{100}{\micro\metre}.

\begin{figure}[t]
    \centering
    \includegraphics[width=0.485\linewidth]{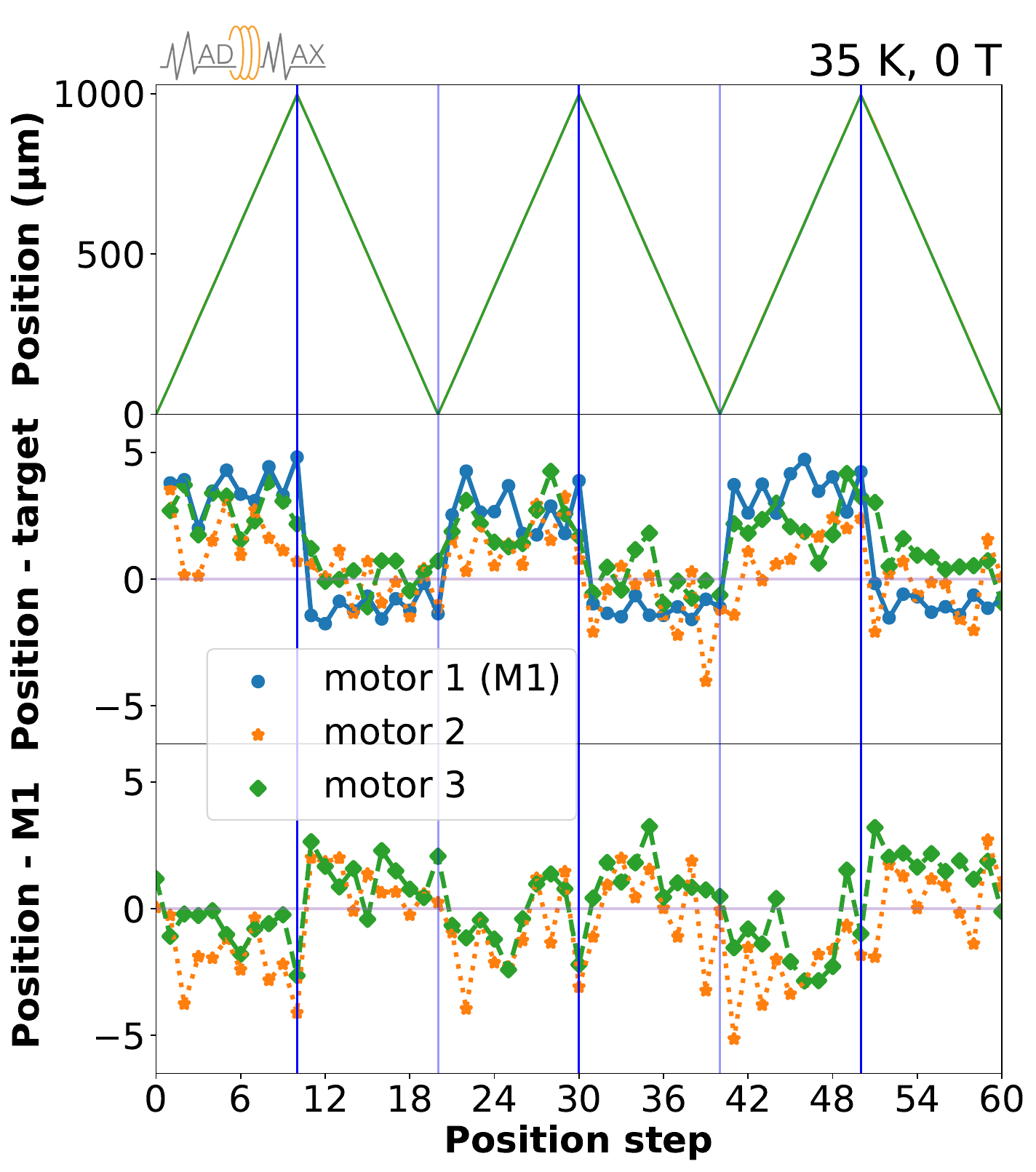}
    \includegraphics[width=0.491\linewidth]{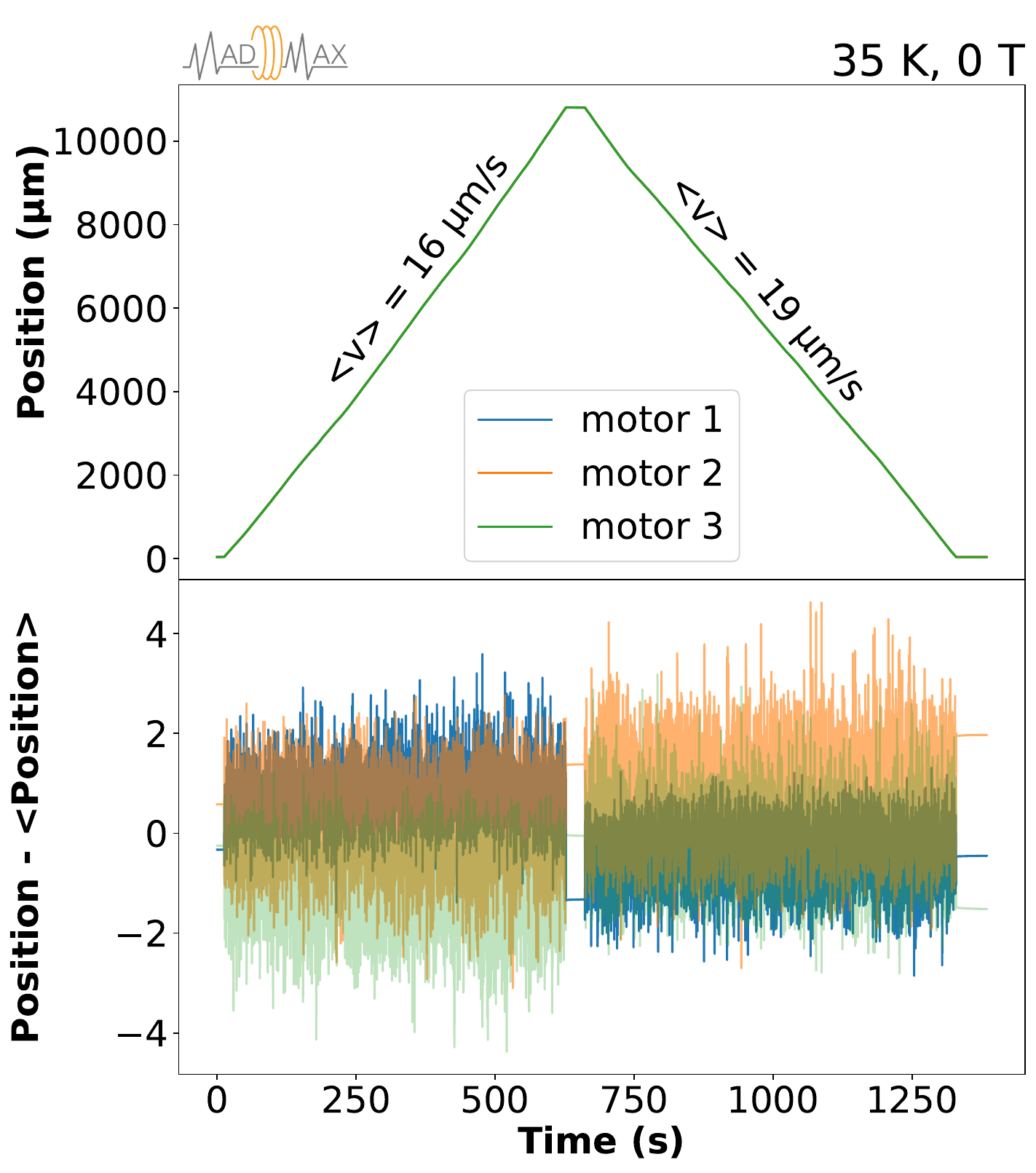}
    \caption{Measurements of OB200 at cryogenic temperature. Left: Positioning accuracy measurements in \SI{}{\micro\metre} for three of the back and forth movements performed, delimited by the vertical blue lines. The evolution as a function of the motor position step (reached after each jump of the motor by \SI{100}{\micro\metre}) from the starting point is shown in the top panel. The distance between the positions of the three motors and the target is shown in the central panel and the distance between motors~2 and 3 and motor~1 in the bottom panel. In both cases, the errors on each point are lower than 1~nm and (not visible on the plot) and lines are connecting the points to ease the plot readability. Right: Motor displacement in \SI{}{\micro\metre} as a function of time in seconds for a full back and forth across OB200 is shown in the top plot. Deviation between each motor and the average position of the three motors as a function of time is shown in the bottom panel. Measurements are taken every \SI{0.11}{\second}.}
    \label{fig:Detail_Pos_Cryo}
\end{figure}


\subsection{Results at cryogenic temperature}
\label{sec:CT}

After the setup cooled down to liquid helium temperature, the backbone structure reached temperatures around \SI{10}{\kelvin} but the motor carriages only reached approximately \SI{25}{\kelvin} without a disk installed (first run) and approximately \SIrange{30}{40}{\kelvin} with the disk installed (second and third runs). This can be attributed to the large area of the thermally floating disk with a weak thermal connection to the backbone structure. Figure~\ref{fig:Detail_Pos_Cryo} left shows the evolution of the motor positions at cryogenic temperature. The jump length is set to \SI{100}{\micro\metre} and a cycle corresponds to a back and forth displacement of $10\times\SI{0.1}{\milli\metre}$ from the original position (\num{10} jumps per direction). No change is observed between the three back and forth movements which illustrates the stability of the procedure at cryogenic temperature. All motors can reach the target position with a precision better than \SI{5}{\micro\metre}, which is much reduced compared to room temperature measurements (Figure~\ref{fig:Detail_Pos_RT} and~\ref{fig:Detail_Pos_RT_B}), as the stepsize of the piezo motors decreases with temperature -- enabling a more precise positioning. The motor positions when reaching the target is well below the tolerance of \SI{10}{\micro\metre}, corresponding to the MADMAX requirement.  Similarly, the position of motor~1 is always within \SI{3}{\micro\metre} from the other motors, which is also significantly better than for room temperature measurements. Dividing or multiplying the jump length by \num{10} does not change the results.

Another run where the motor positions are registered every \SI{0.11}{\second} is performed. Figure~\ref{fig:Detail_Pos_Cryo} right shows the motor displacement as a function of time (top) and the deviation of each motor from  the average position. Contrary to the measurement performed at room temperature, only minor differences are observed between the back and forth movements: the velocity is similar for the two directions, around \SI[per-mode=symbol]{17}{\micro\metre\per\second}, and the distance between motors is within $\pm\SI{4}{\micro\metre}$ in both directions and therefore well within the requirements. The improvement in the inter-motor distance during movement is due to the decreased motor velocities at cryogenic temperatures. This allows the drive-system to adjust the motor velocities of the secondary motors before a significant inter-motor offset appears.

\subsection{Extrapolation to final MADMAX}

From the disk velocity and position accuracy measurements performed with the prototype to the final MADMAX experiment two main improvements are expected: first, the motors will operate in a thin gaseous helium atmosphere, which provides a sufficient thermal link to the actively cooled parts. For a single motor this method has already been successfully tested in~\cite{PiezoProto}. Here, the usage of a thin helium gas atmosphere for the whole disk drive system was also tested briefly. As expected, the temperatures of the carriages and disk were approaching the temperature of the backbone structure after insertion of the Helium gas. However, this test had to be aborted after a short time, as in addition to the thermal link between carriages/disk and the backbone structure, the gas also enables a thermal link to the thermal shield (at \SI{50}{\kelvin}) and the outer vacuum vessel at room temperature. These limitations will be overcome in the final set-up.
The second improvement will come with respect to the disk velocity which here was limited by the primary motor likely due to a non-ideal pre-stress setting compared to the other motors, featuring a particular small step size at cryogenic temperatures. This motor will be replaced for the next iteration of tests. \\

As already mentioned, the tests described in this paper are performed at a temperature of about \SI{25}{\kelvin} and under a magnetic field of \SI{1.6}{\tesla} significantly below the harsher conditions of the final experiment (\SIrange{4}{5}{\kelvin} and \SI{9}{\tesla}). However, no significant change in performance are expected for the piezo motor and the laser interferometer: the piezo motor were already successfully tested at \SI{5}{\kelvin} and inside a \SI{5.3}{\tesla} magnetic field~\cite{PiezoProto}, and the laser interferometer are in fact operating at \SI{10}{\kelvin} during these tests (as the sensor heads are mounted on the backbone structure). Moreover all the laser interferometer components are passive and purely non-magnetic.

\section{Conclusions}
\label{sec:Conclusions}

A prototype using the novel dielectric haloscope open booster concept developed by the MADMAX collaboration has been built. The booster is composed of one movable  sapphire disk and a mirror of 200~mm diameter each. The disk is moved in very challenging environments -- cryogenic temperature, \SI{35}{\kelvin}, and high magnetic field, \SI{1.6}{\tesla} -- and meet the experiment requirements in terms of velocity  and disk position accuracy. During the disk movement, the inter-motor distance is always less than \SI{20}{\micro\metre} and the stability of the motor position once the target position is reached is always better than \SI{1}{\micro\metre}. This paves the road to the realization of an advanced prototype built with more disks of larger diameter (\SI{30}{\centi\metre}). This prototype is planned to be tested at CERN in a cryostat that will fit the warm bore of the \SI{1.6}{\tesla} Morpurgo magnet to perform searches for axion-like particles.

\acknowledgments

The authors would like to thank the CERN central cryogenic laboratory for the support during all the measurements at CERN and the CERN magnet team for the loan of a dry cryostat with large volume which allowed to make the test at cryogenic temperature. This work is funded/acknowledges support by the Deutsche Forschungsgemeinschaft (DFG, German Research Foundation) under Germany’s Excellence Strategy – EXC 2121 ``Quantum Universe'' – 390833306. This project received funding from BMBF under project number 05H21PARD1. We acknowledge the support of the MADMAX project by the Max Planck Society.

\addcontentsline{toc}{section}{References}
\bibliographystyle{hunsrt}
\bibliography{P200}   

\end{document}